\documentclass[preprint,prb,superscriptaddress]{revtex4-1}

\usepackage{graphicx}

\usepackage{amstext}

\begin{document}

\title{Dissipation in a Crystallization Process}

\author{Sven Dorosz${}^1$, Thomas Voigtmann${}^2$, Tanja Schilling}
\affiliation{Physics and Materials Science Research Unit, Universit\'e du Luxembourg, L-1511 Luxembourg, Luxembourg\\${}^2$Institut f\"ur Materialphysik im Weltraum, Deutsches Zentrum f\"ur Luft- und Raumfahrt (DLR), 51170 K\"oln, Germany, and Department of Physics, Heinrich-Heine-Universit\"at D\"usseldorf, 40225 D\"usseldorf, Germany}

\begin{abstract}
{We discuss the crystallization process from the supersaturated melt in terms of its non-equilibrium properties. In particular, we quantify the amount of heat that is produced irreversibly when a suspension of hard spheres crystallizes. This amount of heat can be interpreted as arising from the resistance of the system against undergoing phase transition. We identify an intrinsic compression rate that separates a quasi-static regime from a regime of rapid crystallization. In the former the disspated heat grows linearly in the compression rate. In the latter the system crystallizes more easily, because new relaxation channels are opened, at the cost of forming a higher fraction of non-equilibrium crystal structures. In analogy to a shear-thinning fluid, the system shows a decreased resistance when it is driven rapidly.}
\end{abstract}

\maketitle

Crystallization from the metastable melt is a non-equilibrium process. Yet, it is usually discussed in terms of quasi-equilibrium concepts such as transition state theory \cite{Kashchiev2003, Oxtoby2009}, which do not account for the fact that any irreversible process of finite duration is inevitably subject to dissipation. Here, we present a numerical approach to assess the amount of energy that is dissipated in a crystallization process, and we discuss the relation between dissipation and external driving.

The most obvious way to characterize the irreversibility of a process is to quantify entropy production. However, to do so directly is unpractical even for very simple model systems. To bypass this problem, we instead evaluate the mechanical work performed on the system by compressing it at a constant rate, and subtract the equilibrium work, which we obtain independently via the equation of state.

As a model system we use hard spheres, the most simple system that shows a liquid-to-crystal transition \cite{Alder1957}. Despite their simplicity hard spheres capture the essential physics of the crystallization process in many atomic systems. In general, at high densities the excluded volume between atoms dominates their dynamical behaviour, because the typical interparticle distances are smaller than the attraction ranges. The attractive forces thus effectively only add up to a flat background that does not influence the dynamics, but merely changes the equilibrium equation of state \cite{Widom1967,Voigtmann2008}. Thus the results from our study should be applicable to a large class of crystallization processes in metallic systems as well as colloids.

To model the crystallization process, we perform computer simulations of hard spheres in the NPT ensemble. The system is prepared in a fluid equilibrium state at constant pressure and then subjected to an increase in pressure with a constant rate $\dot{P}$ for a duration $\tau$. Under these conditions the work $W$ performed on the system is
\[
W=\int_0^{\tau} dt \dot{P} V(t), 
\]
where $V(t)$ is the volume response of the system to the external driving $\dot{P}$.
Assuming the difference in Gibbs free energy $\Delta G$ between the initial and the final state to be known, the dissipated energy of each simulation trajectory is
\[
W_\text{diss}=W-\Delta G\,. 
\]

\begin{figure}
\begin{center}
\includegraphics[width=0.95\linewidth]{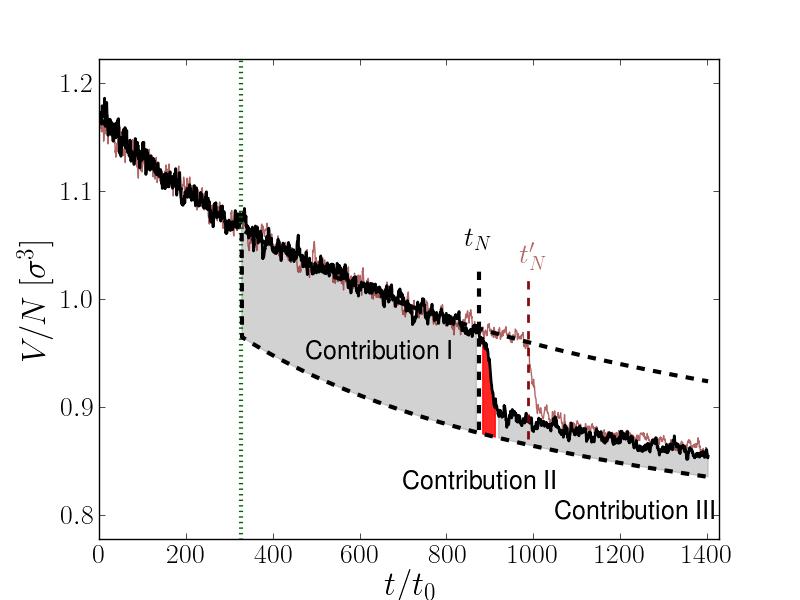}
\end{center}
\caption{\label{fig:trajectory}Evolution of the specific volume along a typical simulation trajectory of a hard-sphere system that is compressed with $\dot P=0.01065\,k_BT/\sigma^3t_0$ (solid lines). Induction times are labeled $t_N$ and $t_N'$. Dashed lines indicate the equations of state of the fluid (upper line) and the ideal equilibrium crystal (lower line), respectively. A vertical line marks the time when the coexistence pressure $P_c$ is crossed. Shaded areas indicate the different contributions to the work performed on the system, as discussed in the main text. Area (II) is the dissipated heat $q^c$.}
\end{figure}

The volume evolution of a typical trajectory is shown as a solid black line in Fig.~\ref{fig:trajectory} (the volume is given in units of the sphere diameter cubed, $\sigma^3$, time in units of the free diffusion time $t_0$ which we specify in the next paragraph). Dashed lines indicate the equations of state of the liquid \cite{Carnahan1969} and the crystal \cite{Alder1968}. The dissipated energy $W_\text{diss}$ is indicated by the shaded areas. It consists of three contributions: (i) is the work associated with compression of the metastable liquid phase until the nucleation event occurs; (iii) is a contribution that arises because the system is not completely transformed into the equilibrium crystal during the simulation time, but contains defects. This contribution results in an almost constant offset in volume with respect to the equation of state of the equilibrium crystal, which does not vary much between trajectories for any given $\dot P$. Contribution (ii) yields the irreversible heat per particle, $q^c$, associated with the crystallization process,
\[ q^c=\frac1N \int_{t_N}^{t_N+\Delta t}dt\,\dot P\,[V-V_\text{eq}(P)]\,. \]
We model the dynamics of the system as stochastic, thus $q^c$ is a fluctuating quantity on the ensemble of trajectories. 

Before we discuss the distribution of $q^c$, we summarize the technical details of the simulation. We perform standard NPT Monte Carlo (MC) simulations with small particle displacements drawn from a flat distribution from the interval $[-\Delta,\Delta]$ with $\Delta=0.065\sigma$. As unit of time we use $t_0=\sigma^2/D_0$, where the free-particle diffusion coefficient is $D_0=\Delta^2/6/\text{MC sweep}\approx7\times10^{-4}\sigma^2/\text{MC sweep}$. To control the pressure, a volume change is attempted once per MC sweep by rescaling the box lengths according to $L_i\mapsto L_i\,\exp[0.0012(r-1/2)]$ where $r$ is a uniform random variable in $]0,1]$ and $i$ labels the Cartesian directions. We allow changes of $L_i$ independently in each direction to accomodate crystals with unit cells of different aspect ratios.
Simulations start from an equilibrium fluid state at pressure $P_0=8\,k_BT/\sigma^3$ and end at $P_\tau=23\,k_BT/\sigma^3$, where $k_BT$ is the thermal energy and $\tau=(P_\tau-P_0)/|\dot P|$ is the duration of the trajectory. The crystal-liquid coexistence pressure is $P_c=11.54\,k_BT/\sigma^3$ \cite{Noya2008}.
We monitor the degree of crystallinity by means of the local $q_6q_6$ bond order parameter \cite{Steinhardt1983,tenWolde1995}. In order to distinguish different crystal structures we analyze the averaged bond order parameters $|q_4|$ and $|q_6|$ \cite{Lechner2008}. 

To compute $q^c$ we need to define the time window of contribution (ii) (see Fig.~\ref{fig:trajectory}). We set the induction time $t_N$ to the time after which the largest crystalline cluster maintains a size of ten or more particles. The end of the process, $t_N+\Delta t$, is set to the time when the overall crystallinity reaches $60\%$. This value is large enough to capture the main contributions of dissipated heat, but still small enough to minimize the influence of periodic boundary conditions.

Since rare trajectories can contribute considerably to the non-equilibrium work distribution, we need to generate a very large number of independent trajectories. As the computational effort is large ($O(10^5)$ trajectories per value of compression rate), we simulate a relatively small system of $N=540$ particles, a system size that, albeit small, still reproduces the nucleation rates correctly. To verify this, we compared the nucleation rates to those obtained from simulations with $N=8000$ and $N=216,000$ particles \cite{Schilling2011}.

We compressed the system for times $\tau=1\times 10^5$, $2\times10^5$, $5\times10^5$, $1 \times 10^6$, $2\times10^6$, $5\times10^6$, and $1 \times 10^7$ MC sweeps (corresponding to compression rates between $\dot P\approx0.214\,k_BT/\sigma^3t_0$ and $\dot P=0.00214\,k_BT/\sigma^3t_0$.) The number of trajectories sampled varied between $70,000$ and $650,000$ depending on the accuracy needed for a given value of $\dot P$. In total this required 90 years of CPU time on 2.2GHz Xeons \cite{Waldorf1976}.

\begin{figure}
\begin{center}
\includegraphics[width=\linewidth]{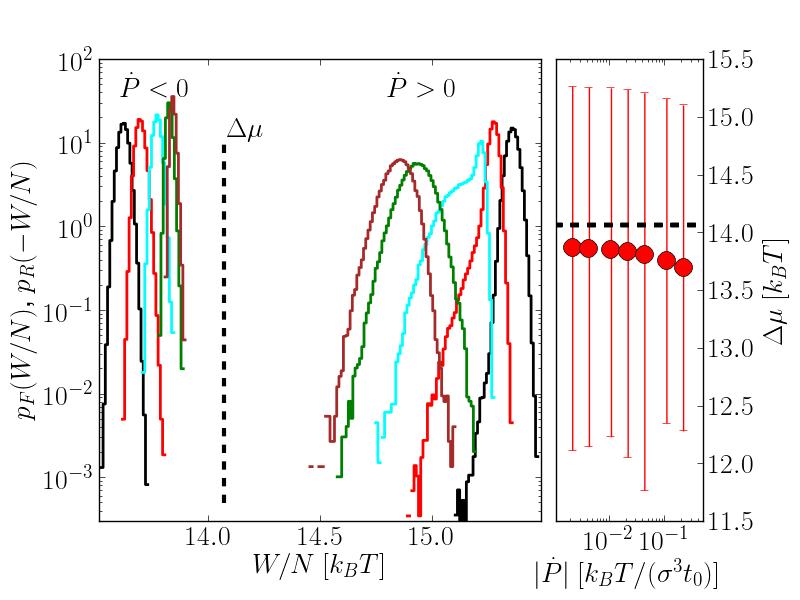}
\end{center}
\caption{\label{fig:jarzynski} Left panel: Distribution of work per particle, $p(W/N)$, performed upon compression (forward process: right set of curves) and expansion (reverse process: left set of curves, shown as $p(-W/N)$) across the phase transition with a constant rate $|\dot P|$. Histograms are shown for $|\dot{P}|=0.214, 0.107, 0.0428, 0.0107, 0.00428\,kT/\sigma^3 t_0$ (right to left for forward process). The dashed vertical line indicates the equilibrium chemical potential difference $\Delta \mu$. Right panel: $\Delta \mu$ estimated using the Jarzynski relation, eq.~\ref{eq:jarzynski}, see text. The horizontal dashed line indicates the equilibrium value.}
\end{figure}

The left panel of Fig.~\ref{fig:jarzynski} shows the distribution of work performed on the ensemble of non-equilibrium trajectories. Solid lines mark $p(W/N)$ for different values of $|\dot P|$ in the forward process ($p_{\rm F}(W/N)$ on compression, $\dot P>0$) and the reverse process ($p_{\rm R}(W/N)$ on expansion shown as $p_{\rm R}(-W/N)$, $\dot P<0$,). The distributions for the expansion processes are centered around values $|W|/N<|\Delta \mu|$, where $\Delta \mu = \Delta G/N$ is the difference in chemical potential between the initial and the final state. For all values of $\dot P < 0$, the curves are well described by Gaussian probability distributions down to the accuracy set by the number of trajectories that we simulated. The distributions $p_{\rm F}(W/N)$ associated to the compression processes are centered around $W/N>\Delta \mu$. In particular at small $\dot P$, they deviate from Gaussian behaviour and display a more subtle structure which we discuss later in terms of $q^c$.

To estimate whether our ensemble averaging is sufficient, we test whether these work distributions are compatible with the Jarzinsky relation \cite{Jarzynski1997} and the Crooks relation \cite{Crooks1999}. For an arbitrary non-equilibrium process, the Jarzinsky relation connects the work distribution $p(W)$ to the equilibrium free energy difference $\Delta G$,
\begin{equation} \langle\exp(-\beta W)\rangle=\int p(W)\exp(-\beta W)\,dW=\exp(-\beta\Delta G)\,. \label{eq:jarzynski} \end{equation}
Since the average is over the exponential of $W$, the Jarzynski relation provides a very sensitive test for the accuracy of sampling.

For distributions $p(W)$ that are superpositions of Gaussian distributions, eq.~\ref{eq:jarzynski} can be evaluated analytically \cite{Hijar2010}. Fitting the work distribution of the forward process with a superposition of two Gaussians, constrained such that the backward
process is also well described by Crooks' fluctuation theorem, we use the distributions $p(W)$ from our simulations to estimate $\Delta\mu$. The 
results are shown in the right panel of Fig.~\ref{fig:jarzynski} (circles with error bars).
Agreement with the equilibrium chemical potential difference obtained from the equation of state is reasonable, thus we conclude that our sampling is 
sufficient.

As discussed above, most of the work $W$ performed on the system during compression consists of equilibrium or quasi-equilibrium contributions that are readily evaluated if one knows the equations of state of the initial and final phase. The non-equilibrium nature of the process is characterized by the distribution of dissipated energy; a quantity that has not been discussed before in the literature on crystallization. 

\begin{figure}
\begin{center}
\includegraphics[width=\linewidth]{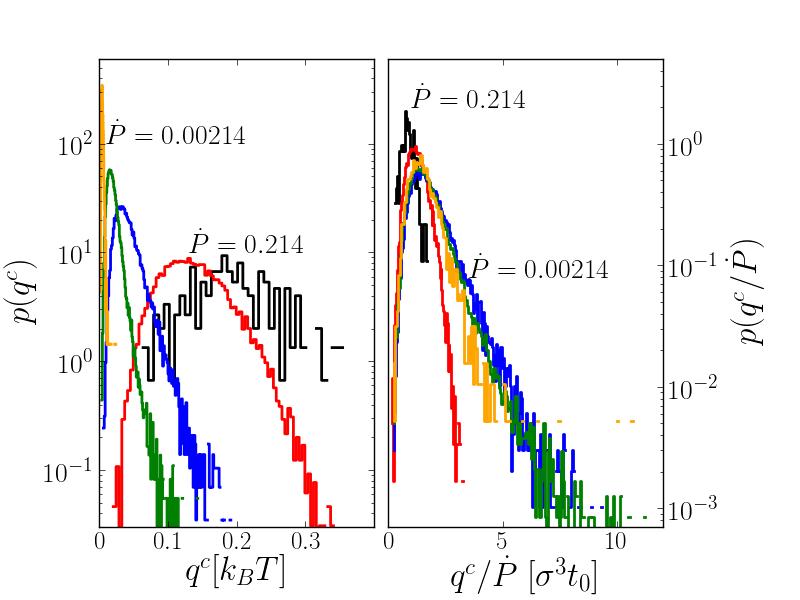}
\end{center}
\caption{ \label{fig:qc_dist}Left panel: Probability distribution of dissipated energy per particle during crystallization, $q^c$, for different compression rates, $\dot P=0.00214, 0.0107, 0.0214, 0.107, 0.214\,kT/\sigma^3 t_0$, as indicated. Right panel: Corresponding distributions of the crystallization loss, $q^c/\dot P$.}
\end{figure}

The left panel of Fig.~\ref{fig:qc_dist} shows the distribution of dissipated energy per particle, $q^c$, for various values of the compression rate $\dot P$. As the compression rate increases, the distribution shifts to higher average $\langle q^c\rangle$ and broadens. At the highest compression rate we simulated, $\langle q^c\rangle$ is about $0.2\,k_BT$, which is of the same order of magnitude as the average (macroscopic) interfacial energy over the area per particle, $\gamma\sigma^2\approx0.6\,k_BT$ \cite{Haertel2012}. In the right panel of the figure we show that the distributions collapse for weak driving $\dot P$, when plotted in terms of the reduced variable $q^c/\dot P$. This collapse defines the regime of quasi-static behaviour, where the response $q^c/\dot P$ of the system is independent of the driving force. 

In the context of equilibrium thermodynamics the term ``quasi-static'' is restricted to the case of infinitely slow driving, $\dot P=0$. The existence of a regime of $\dot P$-independent distributions of $q^c/\dot P$ justifies the extension of this notion to finite (small) driving rates. The limiting value of the average response $\zeta:=\langle q^c\rangle/\dot P$ attained for $\dot P\to0$ can be interpreted as an immanent system property, the \emph{quasi-static crystallization loss}. For the hard-sphere system, we obtain $\zeta_{\dot P\to0}\approx1.6\,\sigma^3t_0$.

\begin{figure}
\begin{center}
\includegraphics[width=0.7\linewidth]{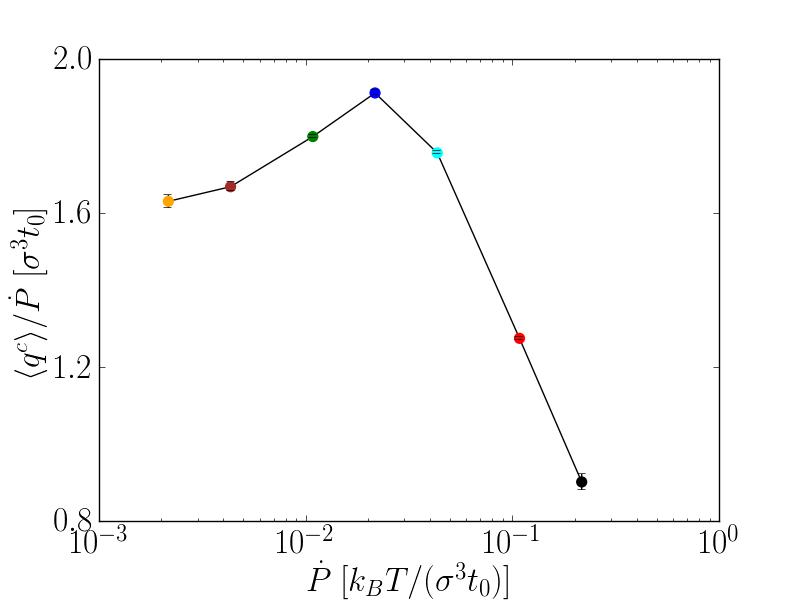}
\end{center}
\caption{\label{fig:central} Crystallization loss, i.e,, the average energy per particle dissipated during crystallization relative to the external driving rate,$\langle q^c\rangle/\dot P$ as a function of compression rate $\dot P$.}

\end{figure}

At driving forces above a threshold $\dot P^*$ the crystallization loss $\zeta(\dot P)$, as a function of $\dot P$, drops sharply (see Fig.~\ref{fig:central}).(If one took into account the energy costs due to defects in the crystal (contribution~(iii) in Fig.~\ref{fig:trajectory}) this effect would even be enhanced, since the excess volume over the equilibrium volume $V_\text{eq}$ (used to define $q^c$) increases monotonically with increasing $\dot P$.)
Intuitively, one would expect the relative dissipation to increase once the rate of driving exceeds typical microscopic relaxation times of the system, as additional work can be dissipated through the microscopic degrees of freedom. The counter-intuitive behaviour of the crystallization loss can be rationalized by analogy with mechanical friction in fluids. There, one typically observes friction to decrease strongly in the nonlinear-response regime of fast driving \cite{Urbakh2004}. This is particularly well known for the viscosity of non-Newtonian fluids \cite{Voigtmann2014}, where the effect is called shear thinning. It also holds for a driven tracer subject to an external force in a dense fluid \cite{Puertas2014}. In these cases, the slow near-equilibrium relaxation processes are replaced by faster ones that occur on the time scale set by the external driving. In analogy, we interpret $\zeta(\dot P)$ as a generalized friction coefficient that characterizes the melt's resistance to phase transformation.

$\zeta(\dot P)$ shows non-monotonic behaviour because both effects, i.e., increased friction through enhanced coupling to microscopic degrees of freedom as well as decreased friction through non-equilibrium relaxation channels, contribute to the crystallization loss. As indicated in Fig.~\ref{fig:qc_dist}, the initial increase is associated with an increase in the large-$q^c$ tail. This is intuitively expected since an increased coupling to microscopic degrees of freedom increases the probability for stronlgy dissipating trajectories. At $\dot P>\dot P^*$, this large-$q^c$ tail is cut off. (We will show in the following that this effect is due to the formation of non-equilibrium crystal structures.) The crossover between the two trends occurs around $\dot P^*\approx2\times10^{-2}\,k_BT/\sigma^3t_0$.This cross-over value is explained by the time scale $t_L$ needed for collective particle rearrangements involving the nearest and next-to-nearest neighbour shells. $t_L$ is set by the long-time self-diffusion coefficient $D_L=\sigma^2/t_L$. For the typical densities reached when crystallization sets in, $D_L/D_0=O(10^{-2})$ (for the initial fluid state in our work, $D_L/D_0\approx0.04$) \cite{Alder1970}. Hence, $\dot P^*t_L=O(k_BT/\sigma^3)$; i.e.~the effects of the external driving start to dominate the crystallization process once the compression rate is faster than the typical thermal energy density can be redistributed through collective particle rearrangements.

\begin{figure}
\begin{center}
\includegraphics[width=0.9\columnwidth]{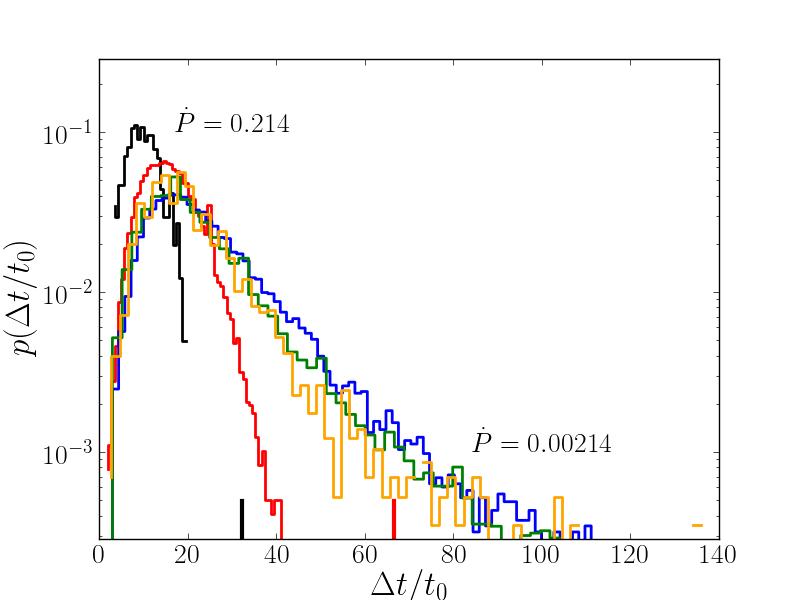}
\end{center}
\caption{ \label{fig:histo_dt}Distribution of the time interval over which the system crystallizes, for different $\dot P$ corresponding to the data shown in Fig.~\ref{fig:qc_dist}. Vertical lines indicate the maximum $\Delta t$ possible in the simulation for the earliest induction time $t_N$ observed in our simulations, for the highest two $\dot P$ shown.}
\end{figure}

To demonstrate that the melt indeed relaxes faster into the crystal phase at $\dot P>\dot P^*$, we show in Fig.~\ref{fig:histo_dt} the distributions of the crystallization time $\Delta t$ (i.e.~the distributions of the length of time between the induction time and the time when 60\% of the system are crystallized). For small $\dot P$, the distributions again collapse to a $\dot P$-independent curve. This curve displays a pronounced tail at large $\Delta t$, and the crystallization process is slow on the time scale $t_0$ of free particle diffusion. The average $\langle\Delta t\rangle$ is approximately $20t_0$, i.e.~on the order of the long-time self-diffusion time $t_L$. This fact confirms that long-time diffusion sets the relevant time scale for the crystallization process. At large $\dot P$, the distributions shift to smaller average $\Delta t$, and the large-$\Delta t$ tail is cut off. Moreover, $p(\Delta t)$ narrows proportionally to $\dot P$, which suggests that the inverse external driving rate $1/\dot P$ sets the relevant time scale for the dynamics. The change in shape of the distribution $p(\Delta t)$ with $\dot P$ is qualitatively similar to the one observed for the distribution $p(q^c/\dot P)$ shown in Fig.~\ref{fig:qc_dist}. This emphasizes that the change in dissipation mechanism is of kinetic rather than thermodynamic origin.

\begin{figure}
\begin{center}
\includegraphics[width=0.9\columnwidth]{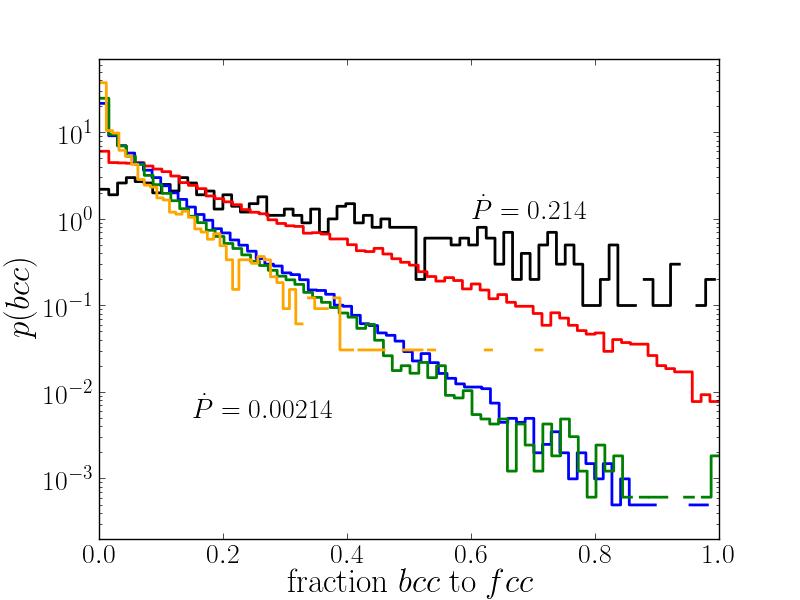}
\end{center}
\caption{ \label{fig:bcc_dist}Distribution of the fraction of bcc crystal structures in the crystalline part of the system at the end of the compression run, for different $\dot P$ corresponding to the data shown in Fig.~\ref{fig:qc_dist}. }
\end{figure}

Next we show that the accelerated crystallization mechanism proceeds through non-equilibrium relaxation channels, in particular the formation of non-equilibrium crystal structures (bcc instead of fcc).
Fig.~\ref{fig:bcc_dist} shows the probability distribution of the fraction of particles with a bcc-like environment in the crystal at the end of the simulation run. Again, at slow driving rates, $\dot P\lesssim\dot P^*$, the distributions are independent of $\dot P$. For $\dot P\gtrsim\dot P^*$, more bcc structures are formed. There even is a signifiant number of runs that crystallize completely into bcc. Our data indicate that the formation of bcc-like structures is facilitated at large compression rates. Interestingly, the question why fcc is the stable equilibrium structure while bcc should form more easily is known from Landau theory \cite{Alexander1978}. There, the effect arises because a larger set of reciprocal lattice vectors is needed to form fcc. This implies that a larger set of local density fluctuations needs to be sampled. It is conceivable that this takes more time, and hence, bcc is favored kinetically.

The tendency to form metastable crystal structures in rapid solidification is well known from metallic melts \cite{Herlach2014}. It is often attributed to Oswald's step rule, which invokes interfaction tensions between the crystal nucleus and the surrounding fluid. We offer an alternative explanation that is founded on microscopic kinetic arguments, rather than macroscopic thermodynamic quantities that might not be well defined on the scale of a few particle diameters.

\section*{Conclusion}
We have discussed crystallization in terms of non-equilibrium notions and calculated the distribution of heat dissipated during a crystallization process. Compressing the system at different rates $\dot P$, we measure the volume response and find two regimes: Below a characteristic compression rate, $\dot P^*$, set by the single-particle diffusion time and the coexistence pressure, the resistance of the system against the phase transition, $\langle q^c\rangle/\dot P$, is constant. The system responds quasi-statically. Above $\dot P^*$ the crystallization process evolves far from equilibrium. The system crystallizes more easily than expected, because new relaxation channels are opened via the formation of bcc structures instead of the thermodynamically favored fcc ones. In this regime the evolution of the system is determined by kinetics rather than thermodynamics.

\begin{acknowledgments}
This project has been financially supported by the National Research Fund, Luxembourg under the project FRPTECD. Data from computer simulations presented in this paper were carried out using the HPC facilities of University of Luxembourg~\cite{VBCG_HPCS14}.
We thank M.~Allen, J.~Horbach, and S.~Williams for useful discussions.
\end{acknowledgments}


\begin{thebibliography}{10}

\bibitem{Kashchiev2003}
D. Kashchiev and G.~M. Van Rosmalen,
\newblock Review: nucleation in solutions revisited.
\newblock {\em Crystal Research and Technology}, 38:555--574, 2003.

\bibitem{Oxtoby2009}
D.~W. Oxtoby
\newblock Nucleation of crystals from the melt.
\newblock {\em Adv. Chem. Phys }, 70:263--296, 2009.

\bibitem{Alder1957} 
B.~J. Alder and T.~E. Wainwright,
\newblock Phase transition for a hard sphere system.
\newblock {\em The Journal of Chemical Physics}, 27(5):1208, 1957.

\bibitem{Widom1967}
B.~Widom,
\newblock Intermolecular Forces and the Nature of the Liquid State.
\newblock {\em Science}, 157:365--382, 1967.

\bibitem{Voigtmann2008}
Th.~Voigtmann,
\newblock Idealized glass transitions under pressure: dynamics versus thermodynamics.
\newblock {\em Phys. Rev. Lett.}, 101:095701, 2008.

\bibitem{Carnahan1969}
Norman~F. Carnahan and Kenneth~E. Starling.
\newblock Equation of state for nonattracting rigid spheres.
\newblock {\em The Journal of Chemical Physics}, 51(2):635--636, 1969.

\bibitem{Alder1968}
B.~J. Alder, W.~G. Hoover, and D.~A. Young.
\newblock Studies in molecular dynamics. v. high‐density equation of state
  and entropy for hard disks and spheres.
\newblock {\em The Journal of Chemical Physics}, 49(8):3688--3696, 1968.

\bibitem{Noya2008}
Eva~G. Noya, Carlos Vega, and Enrique de~Miguel.
\newblock Determination of the melting point of hard spheres from direct
  coexistence simulation methods.
\newblock {\em The Journal of Chemical Physics}, 128(15):--, 2008.

\bibitem{Steinhardt1983}
Paul Steinhardt, David Nelson, and Marco Ronchetti.
\newblock {Bond-orientational order in liquids and glasses}.
\newblock {\em Physical Review B}, 28(2):784--805, 1983.

\bibitem{tenWolde1995}
Pieter~Rein ten Wolde, Maria~J. Ruiz-Montero, and Daan Frenkel.
\newblock Numerical evidence for bcc ordering at the surface of a critical fcc
  nucleus.
\newblock {\em Phys. Rev. Lett.}, 75:2714--2717, Oct 1995.

\bibitem{Lechner2008}
Wolfgang Lechner and Christoph Dellago.
\newblock Accurate determination of crystal structures based on averaged local
  bond order parameters.
\newblock {\em The Journal of Chemical Physics}, 129(11):--, 2008.

\bibitem{Schilling2011}
T.~Schilling, S.~Dorosz, H.~J.~Sch\"ope, and G.~Opletal.
\newblock Crystallization in suspensions of hard spheres: a Monte Carlo and molecular dynamics simulation study.
\newblock {\em J.~Phys.: Condens. Matter}, 23:194120, 2011.

\bibitem{Waldorf1976}
Waldorf and Statler,
\newblock You gotta give them credit.
\newblock {\em The Muppet Show: Avery Schreiber}  1.16, 1976

\bibitem{Jarzynski1997}
C. Jarzynski
\newblock Nonequilibrium Equality for Free Energy Differences.
\newblock {\em Phys. Rev. Lett.}, 78:14, 1977.

\bibitem{Crooks1999} 
G.~E. Crooks, 
\newblock{Phys. Rev. E}, 60:2721 (1999)

\bibitem{Hijar2010}
H.~H\'\i{}jar and J.~M.~Ortiz de Z\'arate,
\newblock Jarzynski's equality illustrated by simple examples.
\newblock {\em Eur. J. Phys.}, 31:1097--1106, 2010.

\bibitem{Haertel2012}
A.~H\"artel, M.~Oettel, R.~E. Rozas, S.~U. Egelhaaf, J.~Horbach, and
  H.~L\"owen.
\newblock Tension and stiffness of the hard sphere crystal-fluid interface.
\newblock {\em Phys. Rev. Lett.}, 108:226101, May 2012.


\bibitem{Urbakh2004}
M.~Urbakh, J.~Klafter, D.~Gourdon, and J.~Israelachvili,
\newblock The nonlinear nature of friction.
\newblock {\em Nature}, 430:525--528, 2004.

\bibitem{Voigtmann2014}
Th.~Voigtmann,
\newblock Nonlinear glassy rheology.
\newblock {\em Curr. Opin. Colloid Interf. Sci.}, 19:549--560, 2014.

\bibitem{Puertas2014}
A.~M.~Puertas and Th.~Voigtmann,
\newblock Microrheology of colloidal systems.
\newblock {\em J.~Phys.: Condens. Matter}, 26:243101, 2014.

\bibitem{Alder1970}
B.~J. Alder, D.~M. Gass, and T.~E. Wainwright,
\newblock Studies in Molecular Dynamics. VIII. The Transport Coefficients for a Hard-Spheres Fluid.
\newblock J. Chem. Phys., 53, 3813--3826, 1970.



\bibitem{Alexander1978}
S. Alexander, J. Mctague,
\newblock
Should All Crystals Be bcc? Landau Theory of Solidification and Crystal Nucleation.
\newblock {\em Phys. Rev. Lett.}, 41:702 (1978)

\bibitem{Herlach2014}
D.~M.~Herlach,
\newblock Non-Equilibrium Solidification of Undercooled Metallic Melts.
\newblock {\em Metals}, 4:196--234, 2014.



\bibitem{VBCG_HPCS14}
S.~Varrette, P.~Bouvry, H.~Cartiaux, and F.~Georgatos.
\newblock {Management of an Academic HPC Cluster: The UL Experience}.
\newblock In {\em Proc. of the 2014 Intl. Conf. on High Performance Computing
  \& Simulation (HPCS 2014)}, Bologna, Italy, July 2014. IEEE.

\end{thebibliography}
\end{document}